\documentclass[12pt]{article}
\usepackage{pic04}
\usepackage{hyperref}
\usepackage{url}
\usepackage{graphicx}

\newcommand{\ppbar}{p \bar{p}}
\newcommand{\ffee}{f \bar{f} \rightarrow e^{\rm +}e^{\rm -}}
\newcommand{\afb}{A_{FB}}

\begin{document}

\title{\bf MEASUREMENT OF Z-QUARK, Z-ELECTRON COUPLINGS \\
AND $\mathbf{\sin^2 \theta_{W}}$ AT CDF}
\author{
G. De Lentdecker, J. Lee, K. McFarland           \\
{\em University of Rochester, NY 14627-0171 USA} \\
A. Gibson, G. Veramendi                          \\
{\em UC Berkeley and LBNL}                       \\
Y-K. Kim                                         \\
{\em University of Chicago / Berkeley}            \\
{\bf on behalf of the CDF Collaboration}}   \maketitle

\baselineskip=14.5pt
\begin{abstract}
The forward-backward charge asymmetry ($A_{FB}$) of
$\ppbar \rightarrow e^{+}e^{-} X$ is sensitive to the vector and axial-vector
couplings of the initial state quarks and final state leptons.
Based on the measurement in $72\ pb^{-1}$ of CDF Run II data,
we measure the coupling constants of the $Z$ boson to
quarks and electron. The electroweak mixing angle, $\sin^2\theta_{W}$,
is also measured through its relation to the coupling constants.
\end{abstract}

\baselineskip=17pt

\section{Introduction}
This study is based on the measurement of the forward-backward charge 
asymmetry of electron-positron pairs at CDF in $\ppbar$ collisions 
with $\sqrt{s}$~=~1.96 TeV at Fermilab~\cite{Greg}. The process
$\ppbar \rightarrow l^{\rm +}l^{\rm -}$, where $l$ is an isolated high 
P$_T$ electron or muon, is mediated primarily by virtual photon at low 
energy~\cite{DY}, and by the $Z$ at $M_{l^{\rm +}l^{\rm -}}$~=~$M_Z$.
Elsewhere the interference of these two processes is an important effect.
The vector and axial-vector couplings of the 
gauge bosons to fermions give rise to an asymmetry in the polar angle of
the lepton momentum in the rest frame of the lepton pair with respect 
to the proton direction.
The annihilation process $f \bar{f} \rightarrow e^{\rm +}e^{\rm -}$ depends
on the helicities of the initial fermion $f$ and the 
final electron $e^{\rm -}$. The amplitude is given by~\cite{RosnerI}:
\begin{equation}
A_{ij} = A(f_i \bar{f} \rightarrow e^{\rm +}e^{\rm -}_j) 
     = - Qe^2 + \frac{\hat{s}}{\hat{s}-M^2_Z+iM_Z \Gamma_Z}C^Z_i(f)C^Z_j(e)\rm{,}
\label{eq:amplitude}
\end{equation} 
where $\hat{s}$ denotes the square of the center of mass energy, and the 
coefficients are given in Table~\ref{tab:rosner_couplings} for 
($i$,$j$) = ($L$,$R$). 
The differential cross section for $\ffee$ is% then
\begin{equation}
\frac{d\hat{\sigma}(\ffee)}{d\cos\theta^*} = \frac{\pi \alpha^2}{8\hat{s}}[(|A_{LL}|^2 + |A_{RR}|^2)(1+\cos\theta^*)^2\\
+ (|A_{LR}|^2 + |A_{RL}|^2)(1-\cos\theta^*)^2 ].
\label{eq:xsection}
\end{equation}
\begin{table}[t!]
\begin{center}
\begin{tabular}{ccccccc} \hline
    &&&Fermion \\ \hline
&$u$ quark & &$d$ quark && Electron \\ 
Boson & $C_L$ & $C_R$ & $C_L$ & $C_R$ & $C_L$ & $C_R$  \\ 
$\gamma$ & 2$e$/3 & 2$e$/3 & $-e/3$& $-e/3$& $-e$& $-e$ \\ 
$Z$ & $g_Z(-\frac{1}{2}+\frac{2}{3}x)$&$g_Z(\frac{2}{3}x)$& 
  $g_Z(\frac{1}{2}-\frac{1}{3}x)$&$g_Z(-\frac{1}{3}x)$ &
      $g_Z(\frac{1}{2}-x)$&$g_Z(-x)$ \\ \hline
\end{tabular}
\caption{Left- and right-handed couplings, where $g_Z \equiv
\frac{e^2}{x(1-x)}$ and $x \equiv \sin^2 \theta_W$.}
\label{tab:rosner_couplings}
\end{center}
\end{table}
The forward-backward asymmetry is directly calculated from the cross section.
\begin{equation}
A_{FB} \equiv \frac{\sigma_F - \sigma_B}{\sigma_F + \sigma_B}
= \frac{3}{4}\frac{|A_{LL}|^2 + |A_{RR}|^2-|A_{LR}|^2-|A_{RL}|^2}
                  {|A_{LL}|^2 + |A_{RR}|^2+|A_{LR}|^2+|A_{RL}|^2}
\rm{.}
\end{equation}
The forward-backward asymmetry ($\afb$) is therefore a direct probe of
the relative strength of the vector and axial-vector couplings between
the $Z$ and the leptons.
The variations of the contributing helicities with $\hat{s}$ and of the
up and down quark pdfs with $x$ and thus $\hat{s}$ mean that individual
chiral couplings of u and d type quarks can be found from $\afb$ over a
range of $M_{ee}$.

\section{Strategy}
The values of the parameters are found by fitting the masurements of $\afb$
in fifteen dielectron invariant mass
regions using 72 $pb^{-1}$ of CDF Run II data~\cite{Greg}.
The theoretical calculation
uses ZGRAD~\cite{ZGRAD}, a Monte Carlo cross section
calculation program for Drell-Yan dielectron and dimuon production
with ${\cal O}(\alpha)$ electroweak corrections.

To fit for Z-quark coupling constants, the couplings between
Z and $\mathsf{u_L}$, $\mathsf{u_R}$, $\mathsf{d_L}$, and $\mathsf{d_R}$
were allowed to change, while all the other parameters in ZGRAD were
fixed to the Standard Model values. Table~\ref{tab:q}
shows the quark coupling values extracted from the best fit, together with
the results from other experiments.
\begin{table}[htb]
\begin{center}
\begin{tabular}{|c|c|c|c|c|} \hline
(1)&      & CDF $\pm$ stat $\pm$ sys  &  PDG                       & SM prediction \\ \cline{2-5}
   &u$_L$ & 0.413$\pm$0.141$\pm$0.061 &  0.330$\pm$0.016           & 0.3459$\pm$0.0002 \\ \cline{2-5}
   &u$_R$ & 0.006$\pm$0.117$\pm$0.050 & -0.176$^{+0.011}_{-0.006}$ & 0.1550$\pm$0.0001 \\ \cline{2-5}
   &d$_L$ &-0.318$\pm$0.204$\pm$0.039 & -0.439$\pm$0.011           & 0.4291$\pm$0.0002 \\ \cline{2-5}
   &d$_R$ &-0.022$\pm$0.370$\pm$0.069 & -0.023$^{+0.070}_{-0.047}$ & 0.0776 \\ \hline \hline
(2)&      & CDF $\pm$ stat $\pm$ sys  &  SLD + LEP                 & SM prediction \\ \cline{2-5}
   &e$_V$ & -0.056$\pm$0.012$\pm$0.013 & -0.03816$\pm$0.00047      & -0.0397$\pm$0.0003 \\ \cline{2-5}
   &e$_A$ & -0.536$\pm$0.122$\pm$0.145 & -0.50111$\pm$0.00035      & -0.5064$\pm$0.0001 \\ \hline \hline
(3)&      & CDF $\pm$ stat $\pm$ sys  &  NuTeV~\cite{Erler}        & LEP EWWG \\ \cline{2-5}
   &$\sin^2\theta_W$& 0.2238$\pm$0.0046$\pm$0.0020 & 0.2364$\pm$0.0016 & 0.23143$\pm$0.00015 \\ \hline
\end{tabular}
\caption{Measurements of (1) Left- and right-handed quark couplings with $Z$,
(2) Left- and right-handed electron couplings with $Z$, (3) effective
$\sin^2\theta_W$.}
\label{tab:q}
\end{center}
\end{table}
The dominant source of the systematic error is the
energy resolution of CDF electromagnetic calorimeter, which is
measured using the rate of $e^{+}e^{-}$ as a function of $M_{ee}$.
Other sources of the systematic error include the uncertainties in PDF,
the effect of material, energy scale and background.
The coupling constants of the $Z$ to left- and right-handed electrons are
measured assuming Standard Model quark couplings and are also shown
in Table~\ref{tab:q}.
Finally, effective $\sin^2 \theta_{W}$ is measured
{\em via} its effects on the quark and electron couplings
as in Table~\ref{tab:q}. The sensitivity is dominated by the
$Z$-electron couplings near the $Z$ pole.

\section{Conclusion}
The value of $\sin^2\theta_{W}$ and the coupling constants
of quarks and electron to the $Z$ boson were measured using 72 $pb^{-1}$ of
CDF Run II data. The result shows agreement with Standard Model.
Since the errors are dominated by the statistical errors,
sensitivity is expected to improve with higher luminosity, and
systematic errors will improve with increased statistics.

\end{document}